\documentclass[11pt,oneside,letterpaper]{article}
\pdfoutput=1

\usepackage{setspace}
\usepackage{graphicx}
\usepackage{subcaption}
\usepackage{amsmath}
\usepackage{amssymb}
\usepackage{physics}
\usepackage{xcolor}
\usepackage{color}
    \definecolor{darkgreen}{rgb}{0,0.5,0}
    \definecolor{darkred}{rgb}{0.5,0,0}
    \definecolor{darkblue}{rgb}{0,0,0.6}
    \definecolor{purple}{rgb}{0.4,.2,0.7}
    \definecolor{black}{rgb}{0,0,0}
\usepackage[hyperfootnotes = true, colorlinks = true, linkcolor = darkred, citecolor = darkred, urlcolor = darkred]{hyperref}
\usepackage{comment}
\usepackage{csquotes}
\usepackage{cite}
\usepackage{float}
\usepackage{units}
\usepackage{tikz}
\usepackage{cancel}
\usepackage{mathtools}
\usepackage{pdfsync}
\usepackage{slashed}
\usepackage{url}
\usepackage{upgreek}
\usepackage{float}
\usepackage{units}
\usepackage{tikz}
\usepackage{tikz-feynman}
\usepackage{cancel}
\usepackage{orcidlink}
\usepackage[margin = 2.2cm]{geometry}
\usepackage[ragged]{footmisc}
\include{feynmandefn}

\newcommand{\be}{\begin{eqnarray}\displaystyle}
\newcommand{\ee}{\end{eqnarray}}
\newcommand{\nn}{\nonumber}

\begin{document}

\thispagestyle{empty}
\begin{center}
    ~\vspace{5mm}
    
    {\Large \bf 
        Soft Theorems in Chern-Simons Matter Theories
    }
    
    \vspace{0.4in}

    {\bf Avi Wadhwa$^1$\orcidlink{0000-0002-6724-0872}}

    \vspace{0.4in}
    
    $^1$ International Centre for Theoretical Sciences, Tata Institute of Fundamental Research,\\Shivakote,  Hesaraghatta Hobli,
    Bengaluru 560089, Karnataka,  India.
    
    \vspace{0.1in}
    
    {$^1$ \tt avi.wadhwa@icts.res.in}
    
    \vspace{0.4in}
    
\end{center}

\begin{abstract}

We describe tree-level soft theorems for Chern-Simons QED and QCD in 4+1 dimensions. Soft theorems have been studied in various context in the past. The universal behavior of the leading and subleading soft theorems is understood as a consequence of the local on-shell gauge invariance of the scattering amplitudes. Chern-Simons terms are gauge-invariant only up to a boundary term. Hence, they make for an interesting case to study the relationship between gauge invariance and soft theorems. In this paper, we show that the subleading soft theorems in QED and QCD get corrected when Chern-Simons terms are introduced and also derive those corrections. We start with a discussion on soft theorems as a consequence of gauge invariance. Then we move on to derive the Chern-Simons vertices. Finally, we calculate the correction to soft theorems due to the addition of Chern-Simons terms for both QED and QCD.

\end{abstract}

\pagebreak

\tableofcontents

\pagebreak

\section*{Notations and Conventions}

Throughout this paper, we work in natural units with $\hbar=c=1$.
We also use the ``mostly minus'' signature $\left(+---\right)$
for the space-time metric. $\epsilon^{\mu\nu\rho\sigma}$ refers to
the Levi-Civita symbol normalized as $\epsilon^{0123}=1$. $\mu$,
$\nu$ denote the Lorentz indices while $\alpha$, $\beta$ denote
the indices of the representation of the gauge group. The latter is
often omitted to tidy up the equations, in which case the order
of multiplication is important. Lorentz indices are raised and lowered
with the Minkowski metric $\eta_{\mu\nu}$. Whenever mentioned, the
$\gamma^{\mu}$ denote the Dirac matrices satisfying the Clifford
algebra $\left\{ \gamma^{\mu},\gamma^{\nu}\right\} =2\eta^{\mu\nu}$.
We use the Feynman slash notation where $\cancel{b}=b^{\mu}\gamma_{\mu}$
for any four vector $b^{\mu}$. $A_{\mu}$, $a_{\mu}$ are spin-1
gauge-fields in Lorentz gauge with the latter (with small $a$) denoting
a soft gauge mode. $F_{\mu\nu}$ is the field strength tensor. We work with SU($N$) gauge theories for which
$a_{\mu}$ etc. are matrices in the representation. They can be expanded
in terms of the generators of the gauge group $a_{\mu}\left(x\right)=a_{a\mu}\left(x\right)T_{a}$.
The following types of shorthand for index contractions will be present,
\[
CD:=C_{\alpha}D^{\alpha},
\]
\[
C.D:=C_{\mu}D^{\mu}.
\]
We also use the superscript $T$ to denote the matrix transpose in
the representation indices $\alpha$, $\beta$. Einstein summation
convention is used for all index contractions. Therefore,
\[
C_{a}D_{a}\equiv\sum_{a}C_{a}D_{a}.
\]

\section{Introduction} \label{Intro}

Soft theorems are statements about scattering amplitudes of processes
involving external states with low energy (soft) gauge bosons. They relate the amplitudes with external soft gauge bosons to those
without them via a soft operator. This soft operator is given as an expansion in the soft momenta of the external gauge bosons. The leading term in the expansion has a pole divergence
in the soft momenta \cite{Gell-Mann:1954wra,Low:1954kd,Low:1958sn,Burnett:1967km,Weinberg:1964ew,Weinberg:1965nx}. The leading and subleading terms in the soft operator have a universal form which is completely insensitive to the details of the process and only depends on the charge, momenta and angular momentum of the hard particles. This universal behavior is understood to be a consequence of the local on-shell gauge invariance of the scattering amplitudes  \cite{Low:1954kd,Low:1958sn,Weinberg:1964ew,Burnett:1967km,Bern:2014vva}. This naturally
leads to the following question: does this statement apply to gauge theories with Chern-Simons terms as well? Unlike the usual gauge-invariant terms in the QED or QCD action, the Chern-Simons terms are gauge-invariant only up to a boundary term \cite{Chern:1974ft,Dunne:1998qy}. Therefore, they do not modify the Ward identity corresponding to small gauge transformations. Therefore, the gauge invariance of these terms is more subtle. Naturally then it is interesting to ask whether they modify the universal form of the leading or subleading soft theorems. In this
paper, we answer this question. We calculate how the soft theorems
get corrected at the tree-level when we introduce a Chern-Simons term in QED and QCD.

Chern-Simons terms are only defined in odd space-time dimensions.
They are usually studied in 2+1 space-time dimensions. We are instead working in 4+1 dimensions because we want to treat the Chern-Simons term
perturbatively about the usual gauge theory of QED or QCD.\footnote{In D=2+1, the Chern-Simons term goes like $\sim\mathbb{A}d\mathbb{A}+\mathbb{A}^{3}$
which dominates over the Yang-Mills action in the infrared. This not
the case in higher dimensions where the Chern-Simons term becomes
less relevant in the infrared.}  
Hence, we consider a gauge field coupled to matter in $D=4+1$,
\begin{equation}
S_{\text{gauge}}=\int d^{4+1}x\left\{ -\frac{1}{4g^{2}}F^{2}-\frac{1}{2g^{2}}\left(\partial.A\right)^{2}+\mathcal{L}_{\text{matter}}+\mathcal{L}_{\text{int}}\right\},
\end{equation}
where $A_{\mu}$ is the gauge field and $F_{\mu\nu}$  is the appropriate field strength tensor.

To this we can then add a deformation of the form,

\begin{equation}
S_{\text{CS}}=\kappa\int\mbox{Tr}\left[\mathbb{A}\left(d\mathbb{A}\right)^{2}+c_{4}\mathbb{A}^{3}d\mathbb{A}+c_{5}\mathbb{A}^{5}\right],
\label{non-Abelian CS}
\end{equation}

In the case of QED, the gauge field is Abelian and hence the last two terms vanish. The first term by itself is then the Abelian Chern-Simons term. For QCD, the deformation becomes gauge-invariant (up to a boundary term) when $\left\{ c_{4},c_{5}\right\} =\left\{ \frac{3}{2},\frac{3}{5}\right\} $. Then 
$S_{\text{CS}}$ becomes the non-Abelian Chern-Simons term.\footnote{For non-Abelian Chern-Simons term, under large gauge transformations the action is also shifted by a
constant which is referred to as the winding number \cite{Dunne:1998qy}.
This does not affect the path integral since for appropriate discrete values
of $\kappa$, the winding number is $2\pi$ times an integer and hence
leaves $e^{\mathrm{i}S}$ invariant.}

The total action is then given by the sum of the two,
\[
S=S_{\text{gauge}}+S_{\text{CS}},
\]

In the absence of the term $S_{\text{CS}}$, the subleading
soft theorem (for instance, for an external soft photon with momentum $k_{\mu}$
and polarization $e_{\mu}$) takes the form, 
\begin{equation}
G_{n+1}\left(e,k;\varepsilon_{i},p_{i}\right)=\sum_{i}\varepsilon_{i}^{T}q_{i}\frac{e.p_{i}}{p_{i}.k}G_{n}^{\left(i\right)}\left(p_{i}\right)-\mathrm{i}\sum_{i}\varepsilon_{i}^{T}q_{i}\frac{e_{\mu}k_{\nu}}{p_{i}.k}\left(\hat{J}_{\left(i\right)}^{\mu\nu}G_{n}^{\left(i\right)}\right)\left(p_{i}\right),\label{soft-theorem}
\end{equation}
where $\varepsilon_{i},p_{i}$ refers to the polarization and energy
of the external finite energy (``hard'') particles and $q_{i}$
refers to the charge of the matter coupling to the photon field. Similar
expressions are there for gluons and gravitons as well. The leading
term has a $\nicefrac{1}{k}$ pole divergence while the subleading term
is $\mathcal{O}\left(k^{0}\right)$. The soft photon theorem at leading order in $k$ does not get infrared corrections at loop level \cite{Bern:2014vva, Bern:2014oka, He:2014bga, He:2014cra}. In $D=2+1$, the subleading term gets corrected due to infrared effects at loop level, but in $D=4+1$ the subleading soft photon
theorem is also valid to all loop orders.
The soft gluon theorem in $D=2+1$ admits infrared corrections at loop level even
at the leading order but in $D=4+1$, the theory becomes IR finite and corrections occur only at subleading order. In the case of multiple soft gauge bosons,
each gauge boson brings a soft factor of divergence $\nicefrac{1}{k}$ at
leading order.  

As mentioned before, the leading and subleading behavior in (\ref{soft-theorem}) is universal and can be understood as a consequence of local on-shell gauge invariance of the scattering amplitudes. This was shown for photons in \cite{Low:1958sn}  and for gluons in \cite{Bern:2014vva, Broedel:2014fsa}. For gravitons, even the second subleading behaviour (not stated here) is universal and can be obtained from gauge invariance of the amplitudes \cite{Weinberg:1964ew,Weinberg:1965nx,Bern:2014vva, Broedel:2014fsa}. Sen's covaraintisation procedure \cite{Sen:2017nim,Sen:2017xjn,AtulBhatkar:2018kfi,Laddha:2017ygw, Chakrabarti:2017ltl} utilizes gauge invariance to derive this universal behavior by giving a covariantisation prescription to calculate the 3-point vertex (see $\Gamma^{\left(3\right)}\left(e,k;p,-p-k\right)$ in figure \ref{fig:A1}) for the leading order diagram. Here, one writes an action involving only hard particle fields and covariantises it by minimally coupling the gauge field mode for the soft gauge boson. This is described in detail in section \ref{sen's covariantisation}. Such a covariantisation prescription does not apply to Chern-Simons terms because even though the small gauge Ward identity is still valid, the Chern-Simons terms are not minimally coupled to the matter. Hence, their contribution needs to be accounted for separately in the 3-point vertex, $\Gamma^{\left(3\right)}\left(e,k;p,-p-k\right)$.

Furthermore, we show that gauge invariance is a sufficient but not necessary condition for having the universal form of the leading soft theorems. For this, we will take $c_4$ and $c_5$ in (\ref{non-Abelian CS}) to be arbitrary (and hence make $S_{\text{CS}}$  non-gauge-invariant) and show that the universal form of the leading soft theorem still holds and only the subleading term gets corrected.

Hence, ultimately the aim of this paper is to calculate how (\ref{soft-theorem})
gets corrected at tree-level when a Chern-Simons term  $S_{\text{CS}}$ is added to the gauge theory. 

The paper is organized as follows - in section \S\ref{sen's covariantisation}, we review the
soft photon and gluon theorems and derive them by covariantising the
hard particle action by minimally coupling the soft gauge field modes.
\cite{AtulBhatkar:2018kfi,Sen:2017nim,Chakrabarti:2017ltl,Sen:2017xjn,Laddha:2017ygw}
demonstrate it for soft photons and gravitons. We generalize it to
a generic non-Abelian Yang-Mills theory with SU($N$) gauge invariance
and derive the soft gluon theorem. Soft gluon theorems have been previously
studied in various ways in \cite{Bern:2014vva,Mao:2017tey,Broedel:2014fsa,Broedel:2014bza,Klose:2015xoa}.
Then in section \S\ref{CS vertices}, we calculate the Feynman rules for the Chern-Simons
vertices in $D=4+1$. In section \S\ref{CS correction QED}, \S\ref{CS correction QCD} and \S\ref{CS correction multiple} we incorporate them to obtain
the Chern-Simons correction to the soft theorems. We see that
the leading soft theorems remain unchanged and the Chern-Simons terms
only correct the subleading soft theorems.

\section{Subleading Soft Theorem from Sen's Covariantisation Procedure} \label{sen's covariantisation}

In this section we will summarize the subleading single soft theorem
for photons, as given in \cite{AtulBhatkar:2018kfi} using Sen's covariantisation
procedure \cite{Sen:2017nim,AtulBhatkar:2018kfi,Chakrabarti:2017ltl,Sen:2017xjn,Laddha:2017ygw}
and generalize it for gluons. For gluons, the amplitudes and the corresponding diagrams will be implicitly assumed to be color ordered. At loop level, the familiar soft theorems (\ref{soft-theorem})
get subleading infrared corrections in case of photons and leading
infrared corrections in case of gluons. Therefore, we will restrict
ourselves to only tree-level amplitudes even in this general covariantisation
procedure. We start by considering a scattering process for hard particles,
with amplitude $G_{n}\left(p_{i}\right)$. We assume that the hard
particles in the process are described by a single set of fields $\phi_{\alpha}$,
belonging to some reducible representation of the Lorentz group. Let
$S_{\text{hard}}\left[\phi\right]$ denote the corresponding action
for the hard particle fields. We assume that this action is invariant
under the global transformation,
\[
\phi_{\alpha}\rightarrow\left[e^{\mathrm{i}\theta_{a}T_{a}}\right]_{\alpha}^{\,\,\,\,\beta}\phi_{\beta}.
\]
Here $T_{a}$ are the generators of the Lie group of the gauge symmetry
in the reducible representation and $\theta_{a}$ are constant parameters
for the global symmetry. The generators $T_{a}$ satisfy the Lie algebra,
\[
\left[T_{a},T_{b}\right]=\mathrm{i}f_{abc}T_{c},
\]
where the structure constants $f_{abc}$ are independent of the specific
representation. The algebra is taken to be a priori non-Abelian but
the final results can be rewritten for Abelian fields as a special
case.

\begin{figure}[H]
\centering

\tikzset{every picture/.style={line width=0.75pt}} 

\begin{tikzpicture}[x=0.75pt,y=0.75pt,yscale=-1,xscale=1]

\draw    (83,142) -- (122,142) ;
\draw   (161,142) .. controls (161,128.19) and (172.19,117) .. (186,117) .. controls (199.81,117) and (211,128.19) .. (211,142) .. controls (211,155.81) and (199.81,167) .. (186,167) .. controls (172.19,167) and (161,155.81) .. (161,142) -- cycle ;
\draw   (289,142) .. controls (289,103.06) and (320.56,71.5) .. (359.5,71.5) .. controls (398.44,71.5) and (430,103.06) .. (430,142) .. controls (430,180.94) and (398.44,212.5) .. (359.5,212.5) .. controls (320.56,212.5) and (289,180.94) .. (289,142) -- cycle ;
\draw    (122,142) -- (161,142) ;
\draw [shift={(124,142)}, rotate = 180] [fill={rgb, 255:red, 0; green, 0; blue, 0 }  ][line width=0.08]  [draw opacity=0] (12,-3) -- (0,0) -- (12,3) -- cycle    ;
\draw    (211,142) -- (248,142) ;
\draw [shift={(250,142)}, rotate = 180] [fill={rgb, 255:red, 0; green, 0; blue, 0 }  ][line width=0.08]  [draw opacity=0] (12,-3) -- (0,0) -- (12,3) -- cycle    ;
\draw    (250,142) -- (289,142) ;
\draw    (400,84) -- (433,55) ;
\draw [shift={(431.5,56.32)}, rotate = 318.69] [fill={rgb, 255:red, 0; green, 0; blue, 0 }  ][line width=0.08]  [draw opacity=0] (12,-3) -- (0,0) -- (12,3) -- cycle    ;
\draw    (433,55) -- (464,27) ;
\draw    (429,130) -- (474,117) ;
\draw [shift={(472.08,117.56)}, rotate = 343.89] [fill={rgb, 255:red, 0; green, 0; blue, 0 }  ][line width=0.08]  [draw opacity=0] (12,-3) -- (0,0) -- (12,3) -- cycle    ;
\draw    (474,117) -- (525,103) ;
\draw    (408,194) -- (436,218) ;
\draw [shift={(434.48,216.7)}, rotate = 40.6] [fill={rgb, 255:red, 0; green, 0; blue, 0 }  ][line width=0.08]  [draw opacity=0] (12,-3) -- (0,0) -- (12,3) -- cycle    ;
\draw    (436,218) -- (479,255) ;
\draw  [dash pattern={on 0.84pt off 2.51pt}]  (186,167) -- (186,197) ;
\draw  [dash pattern={on 0.84pt off 2.51pt}]  (186,199) -- (186,239) ;
\draw [shift={(186,197)}, rotate = 90] [fill={rgb, 255:red, 0; green, 0; blue, 0 }  ][line width=0.08]  [draw opacity=0] (12,-3) -- (0,0) -- (12,3) -- cycle    ;
\draw    (429,149) -- (476,164) ;
\draw [shift={(474.09,163.39)}, rotate = 17.7] [fill={rgb, 255:red, 0; green, 0; blue, 0 }  ][line width=0.08]  [draw opacity=0] (12,-3) -- (0,0) -- (12,3) -- cycle    ;
\draw    (476,164) -- (528,179) ;

\draw (162,149) node [anchor=north west][inner sep=0.75pt]   [align=left] {$ $};
\draw (174,134) node [anchor=north west][inner sep=0.75pt]   [align=left] {$\displaystyle \Gamma ^{( 3)}$};
\draw (337,119) node [anchor=north west][inner sep=0.75pt]  [font=\LARGE] [align=left] {$\displaystyle G_{n}^{( i)}$};
\draw (110,116) node [anchor=north west][inner sep=0.75pt]   [align=left] {$\displaystyle \varepsilon _{i}$,$\displaystyle p_{i}$};
\draw (218,115) node [anchor=north west][inner sep=0.75pt]   [align=left] {$\displaystyle p_{i} +k$};
\draw (198,183) node [anchor=north west][inner sep=0.75pt]   [align=left] {$\displaystyle e$,$\displaystyle k$};
\draw (389,37) node [anchor=north west][inner sep=0.75pt]   [align=left] {$\displaystyle \varepsilon _{1}$,$\displaystyle p_{1}$};
\draw (427,76) node [anchor=north west][inner sep=0.75pt]   [align=left] {$\displaystyle ...\varepsilon _{i-1}$,$\displaystyle p_{i-1}$};
\draw (431,178) node [anchor=north west][inner sep=0.75pt]   [align=left] {$\displaystyle \varepsilon _{i+1}$,$\displaystyle p_{i+1}$...};
\draw (397,227) node [anchor=north west][inner sep=0.75pt]   [align=left] {$\displaystyle \varepsilon _{n}$,$\displaystyle p_{n}$};

\end{tikzpicture}
\caption{$A1$: Soft boson attaches to the external leg of a finite energy
particle via a 1PI vertex $\Gamma^{\left(3\right)}$\label{fig:A1}.}
\end{figure}
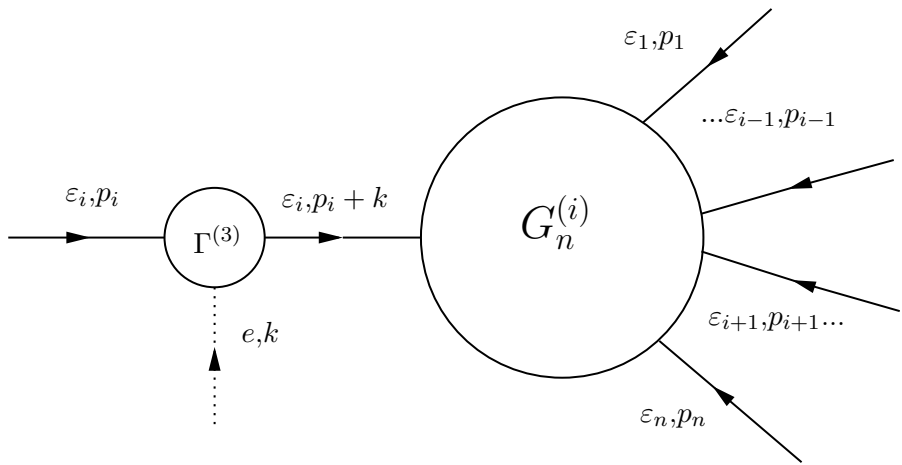

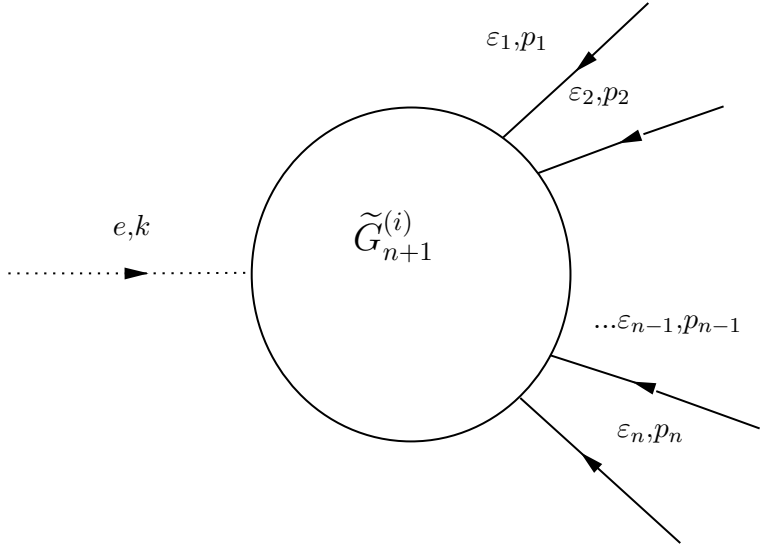
\begin{figure}[H]
\centering
\centering{}

\tikzset{every picture/.style={line width=0.75pt}} 

\begin{tikzpicture}[x=0.75pt,y=0.75pt,yscale=-1,xscale=1]

\draw  [dash pattern={on 0.84pt off 2.51pt}]  (139,172) -- (207,172) ;
\draw   (260.46,172.25) .. controls (260.63,125.87) and (296.38,88.4) .. (340.32,88.56) .. controls (384.26,88.73) and (419.75,126.45) .. (419.58,172.83) .. controls (419.41,219.21) and (383.65,256.68) .. (339.71,256.52) .. controls (295.77,256.36) and (260.29,218.63) .. (260.46,172.25) -- cycle ;
\draw  [dash pattern={on 0.84pt off 2.51pt}]  (207,172) -- (260.02,171.71) ;
\draw [shift={(209,171.99)}, rotate = 179.69] [fill={rgb, 255:red, 0; green, 0; blue, 0 }  ][line width=0.08]  [draw opacity=0] (12,-3) -- (0,0) -- (12,3) -- cycle    ;
\draw    (385.97,103.63) -- (423.33,69.22) ;
\draw [shift={(421.86,70.58)}, rotate = 317.36] [fill={rgb, 255:red, 0; green, 0; blue, 0 }  ][line width=0.08]  [draw opacity=0] (12,-3) -- (0,0) -- (12,3) -- cycle    ;
\draw    (423.33,69.22) -- (458.44,36) ;
\draw    (403.5,121.54) -- (445,106) ;
\draw [shift={(443.13,106.7)}, rotate = 339.48] [fill={rgb, 255:red, 0; green, 0; blue, 0 }  ][line width=0.08]  [draw opacity=0] (12,-3) -- (0,0) -- (12,3) -- cycle    ;
\draw    (456,102) -- (496,88) ;
\draw    (394.52,234.68) -- (426.01,263.38) ;
\draw [shift={(424.54,262.04)}, rotate = 42.35] [fill={rgb, 255:red, 0; green, 0; blue, 0 }  ][line width=0.08]  [draw opacity=0] (12,-3) -- (0,0) -- (12,3) -- cycle    ;
\draw    (426.01,263.38) -- (474.38,307.64) ;
\draw    (409.41,213.17) -- (452,227) ;
\draw [shift={(450.1,226.38)}, rotate = 17.99] [fill={rgb, 255:red, 0; green, 0; blue, 0 }  ][line width=0.08]  [draw opacity=0] (12,-3) -- (0,0) -- (12,3) -- cycle    ;
\draw    (462.39,231.23) -- (514,250) ;

\draw (261.82,181.77) node [anchor=north west][inner sep=0.75pt]  [rotate=-0.21] [align=left] {$ $};
\draw (309.7,138.57) node [anchor=north west][inner sep=0.75pt]  [font=\Large,rotate=-0.26] [align=left] {$\displaystyle \widetilde{G}_{n+1}^{( i)}$};
\draw (189.69,141.26) node [anchor=north west][inner sep=0.75pt]  [rotate=-0.21] [align=left] {$\displaystyle e$,$\displaystyle k$};
\draw (376.19,49.33) node [anchor=north west][inner sep=0.75pt]  [rotate=-0.21] [align=left] {$\displaystyle \varepsilon _{1}$,$\displaystyle p_{1}$};
\draw (417.28,76.95) node [anchor=north west][inner sep=0.75pt]  [rotate=-0.21] [align=left] {$\displaystyle \varepsilon _{2}$,$\displaystyle p_{2}$};
\draw (429.48,191.46) node [anchor=north west][inner sep=0.75pt]  [rotate=-0.21] [align=left] {$\displaystyle ...\varepsilon _{n-1}$,$\displaystyle p_{n-1}$};
\draw (441.46,246.67) node [anchor=north west][inner sep=0.75pt]  [rotate=-0.21] [align=left] {$\displaystyle \varepsilon _{n}$,$\displaystyle p_{n}$};

\end{tikzpicture}
\caption{$A2$: Soft boson attaches to an internal line\label{fig:A2}.}
\end{figure}

To find the amplitude $G_{n+1}\left(e,k;\varepsilon_{i},p_{i}\right)$
for a process with one additional soft boson in the external state,
we need to consider two kinds of Feynman diagrams, $A1$ and $A2$
as shown in figure \ref{fig:A1} and \ref{fig:A2}.
 
Here $\Gamma^{\left(3\right)}$ is the tree-level 3-point vertex involving
a soft boson as well as two hard particles in the external state.
$G_{n}^{\left(i\right),\alpha_{i}}$ is the full $n$-point amplitude
involving $n$ external hard particles with $i$th particle polarization
amputated. To evaluate these diagrams we need to find the vertices.
These can be obtained by taking $S_{\text{hard}}$ and covariantising it
with respect to the single mode soft gauge field $\left[a_{\mu}\left(x\right)\right]_{\alpha}^{\,\,\,\,\beta}=\left[e_{\mu}\left(k\right)\right]_{\alpha}^{\,\,\,\,\,\beta}\exp\left(\mathrm{i}k.x\right)$
to get $S_{\text{soft}}$. This involves replacing all ordinary derivatives
$\partial_{\mu}$ in the action with covariant derivatives $D_{\mu}$
as follows-
\begin{equation}
D_{\mu}\phi_{\alpha}=\partial_{\mu}\phi_{\alpha}-\mathrm{i}\left[a_{\mu}\right]_{\alpha}^{\,\,\,\,\beta}\phi_{\beta},\label{eq: gauge transform}
\end{equation}
as well as all momentum conserving delta functions $\delta\left(p_{1}+\ldots+p_{N}\right)$
by $\delta\left(p_{1}+\ldots+p_{N}+k\right)$. The gauge field $a_{\mu}$
lies in the tangent space of the gauge group and hence can be expanded
as,
\[
\left[a_{\mu}\left(x\right)\right]_{\alpha}^{\,\,\,\,\beta}=a_{a\mu}\left(x\right)\left[T_{a}\right]_{\alpha}^{\,\,\,\,\beta},
\]
and hence,
\[
\left[e_{\mu}\left(k\right)\right]_{\alpha}^{\,\,\,\,\beta}=e_{a\mu}\left(k\right)\left[T_{a}\right]_{\alpha}^{\,\,\,\,\beta}.
\]
We will often omit the representation indices for cleaner looking
equations.

To be more explicit, suppose we want to evaluate $\Gamma^{\left(3\right)}$.
For this, consider the quadratic part of $S_{\text{hard}}$ which can be
written as\footnote{Note that we use $\phi$ to denote the field in both position space
and momentum space.},
\begin{align}
S_{\text{hard}}^{\left(2\right)} & =\frac{1}{2}\int d^{D}xd^{D}y\phi_{\alpha}\left(x\right)\hat{K}^{\alpha\beta}\left(x,y\right)\phi_{\beta}\left(y\right)\label{eq:kinetic-action}\\
 & =\frac{1}{2}\int\frac{d^{D}p_{1}}{\left(2\pi\right)^{D}}\frac{d^{D}p_{2}}{\left(2\pi\right)^{D}}\phi_{\alpha}\left(p_{1}\right)K^{\alpha\beta}\left(p_{2}\right)\phi_{\beta}\left(p_{2}\right)\left(2\pi\right)^{D}\delta^{D}\left(p_{1}+p_{2}\right),
\end{align}
with, 
\[
\hat{K}^{\alpha\beta}\left(x,y\right)=\hat{K}^{\beta\alpha}\left(y,x\right),
\]
\begin{equation}
K^{\alpha\beta}\left(p\right)=K^{\beta\alpha}\left(-p\right),\label{eq:symmetry of K}
\end{equation}
or equivalently,
\begin{equation}
K\left(p\right)=K^{T}\left(-p\right).\label{eq:symmetry of K-1}
\end{equation}
This gives us the Feynman propagator which takes the form,
\begin{equation}
\Delta_{\alpha\beta}\left(p\right)=\mathrm{i}K_{\alpha\beta}^{-1}\left(p\right)=\frac{\Xi_{\alpha\beta}\left(p\right)}{p^{2}-m^{2}},\label{eq:propagator}
\end{equation}
here $\Xi_{\alpha\beta}$ is the polarization sum (times $\mathrm{i}$) for
the hard particle fields. Invariance of (\ref{eq:kinetic-action})
under global transformation (\ref{eq: gauge transform}) implies,
\begin{equation}
T_{a}^{T}K+KT_{a}=0,\label{eq:QK}
\end{equation}
multiplying by $K^{-1}$ on both sides of (\ref{eq:QK}) and using
(\ref{eq:propagator}) gives,
\begin{equation}
T_{a}\Xi+\Xi T_{a}^{T}=0,
\end{equation}
If $p$ is an on shell momentum of an external hard particle of polarization
$\varepsilon\left(p\right)$, then
\begin{equation}
\varepsilon_{\alpha}\left(p\right)K^{\alpha\beta}\left(-p\right)=0.\label{eq:onshell}
\end{equation}
or equivalently
\begin{equation}
\varepsilon^{T}\left(p\right)K\left(-p\right)=0.\label{eq:onshell-1}
\end{equation}
(\ref{eq:propagator}) then implies,
\begin{equation}
\frac{\partial K\left(-p\right)}{\partial p_{\mu}}\Xi\left(-p\right)=2\mathrm{i}p^{\mu}-K\left(-p\right)\frac{\partial\Xi\left(-p\right)}{\partial p_{\mu}},\label{eq:onshell1}
\end{equation}
\begin{equation}
\frac{\partial\Xi\left(-p\right)}{\partial p_{\mu}}K\left(-p\right)=2\mathrm{i}p^{\mu}-\Xi\left(-p\right)\frac{\partial K\left(-p\right)}{\partial p_{\mu}},\label{eq:onshell2}
\end{equation}
for the external hard particle. We have suppressed the representation
indices.

Covariantising of (\ref{eq:kinetic-action}) with respect to a single
soft mode gauge field $a_{\mu}\left(x\right)=e_{\mu}\left(k\right)\exp\left(\mathrm{i}k.x\right)$
using the procedure above gives the following terms in the action
$S_{\text{soft}}$ contributing to the vertex $\Gamma^{\left(3\right)}$
,
\begin{align}
S_{\text{soft}}^{\left(3\right)}= & \frac{1}{2}\int\frac{d^{D}p_{1}}{\left(2\pi\right)^{D}}\frac{d^{D}p_{2}}{\left(2\pi\right)^{D}}\left(2\pi\right)^{D}\delta^{\left(D\right)}\left(p_{1}+p_{2}+k\right)\nonumber \\
 & \phi_{\alpha}\left(p_{1}\right)\left[-\frac{\partial K^{\alpha\gamma}}{\partial p_{2\mu}}\left(p\right)e_{\mu}{}_{\gamma}^{\,\,\,\,\beta}-\frac{1}{2}\frac{\partial^{2}K^{\alpha\gamma}}{\partial p_{2\mu}\partial p_{2\nu}}\left(p\right)k_{\mu}e_{\nu}{}_{\gamma}^{\,\,\,\,\beta}\right]\phi_{\beta}\left(p_{2}\right),\label{eq:cov-kinetic-action}
\end{align}
or equivalently,
\begin{align}
S_{\text{soft}}^{\left(3\right)}= & \frac{1}{2}\int\frac{d^{D}p_{1}}{\left(2\pi\right)^{D}}\frac{d^{D}p_{2}}{\left(2\pi\right)^{D}}\left(2\pi\right)^{D}\delta^{\left(D\right)}\left(p_{1}+p_{2}+k\right)\nonumber \\
 & \phi\left(p_{1}\right)\left[-\frac{\partial K}{\partial p_{2\mu}}\left(p\right)e_{\mu}-\frac{1}{2}\frac{\partial^{2}K}{\partial p_{2\mu}\partial p_{2\nu}}\left(p\right)k_{\mu}e_{\nu}\right]\phi\left(p_{2}\right).\label{eq:cov-kinetic-action-1}
\end{align}
To see how these two terms come, assume that in position space $\hat{K}\left(x,y\right)$
has the general form,
\[
\hat{K}\left(x,y\right)\phi\left(y\right)=\delta^{\left(D\right)}\left(x-y\right)\left[C_{1}+C_{2}^{\mu}\partial_{\mu}+C_{3}^{\mu\nu}\partial_{\mu}\partial_{\nu}\right]\phi\left(y\right),
\]
or equivalently
\[
K\left(p\right)=C_{1}+\mathrm{i}C_{2}^{\mu}p_{\mu}-C_{3}^{\mu\nu}p_{\mu}p_{\nu},
\]
where we have suppressed the $\alpha,\beta$ indices on $\hat{K}$,
$\phi$ and $C_{1,2,3}$. This gives, 
\begin{align*}
\frac{\partial K}{\partial p_{\mu}}\left(p\right)=\mathrm{i}C_{2}^{\mu}-2C_{3}^{\mu\nu}p_{\nu}, & \hspace{1em}\frac{\partial^{2}K}{\partial p_{\mu}\partial p_{\nu}}\left(p\right)=-2C_{3}^{\mu\nu}.
\end{align*}
Then covariantising such a term is achieved by replacing $\partial_{\mu}\phi\rightarrow D_{\mu}\phi$
and $\partial_{\mu}\partial_{\nu}\phi\rightarrow D_{\mu}D_{\nu}\phi$.
In momentum space, the effect of the first is to make the replacement
$q_{\mu}\phi_{\alpha}\rightarrow q_{\mu}\phi_{\alpha}+e_{\mu}{}_{\alpha}^{\,\,\,\,\gamma}\phi_{\gamma}$
which gives the first term in (\ref{eq:cov-kinetic-action}). For
the second term we have,
\begin{align*}
D_{\mu}D_{\nu}\phi_{\alpha}= & \left(\left.\delta_{\alpha}\right.^{\gamma}\partial_{\mu}-\mathrm{i}\,a_{\mu}{}_{\alpha}^{\,\,\,\,\gamma}\right)\left(\left.\delta_{\gamma}\right.^{\beta}\partial_{\nu}-\mathrm{i}\,a_{\nu}{}_{\gamma}^{\,\,\,\,\beta}\right)\phi_{\beta}\\
= & \partial_{\mu}\partial_{\nu}\phi_{\alpha}-\mathrm{i}\,\left(\partial_{\mu}a_{\nu}\right)_{\alpha}^{\,\,\,\,\beta}\phi_{\beta}-\mathrm{i}\,\left(a_{\mu}\partial_{\nu}+a_{\nu}\partial_{\mu}\right)_{\alpha}^{\,\,\,\,\beta}\phi_{\beta}+\mathcal{O}\left(a^{2}\right),
\end{align*}
which in momentum space gives the additional terms,
\begin{align*}
\left[C_{1}+C_{2}^{\mu}D_{\mu}+C_{3}^{\mu\nu}D_{\mu}D_{\nu}\right]\phi & \xRightarrow{\left(\text{momentum space}\right)}\left[K\left(p\right)-\mathrm{i}\,C_{2}^{\mu}e_{\mu}+2C_{3}^{\mu\nu}p_{\mu}e_{\nu}+C_{3}^{\mu\nu}k_{\mu}e_{\nu}\right]\phi\\
= & \left[K\left(p\right)-\frac{\partial K}{\partial p_{\mu}}\left(p\right)e_{\mu}-\frac{1}{2}\frac{\partial^{2}K}{\partial p_{\mu}\partial p_{\nu}}\left(p\right)k_{\mu}e_{\nu}\right]\phi,
\end{align*}
which is same as (\ref{eq:cov-kinetic-action}). These terms in the
action will give the vertex $\Gamma^{\left(3\right)}$ in figure \ref{fig:A1},
which is obtained by symmetrising with respect to the external hard
momenta,
\begin{align*}
 & \Gamma^{\left(3\right)\alpha\beta}\left(e,k;p,-p-k\right)\\
 & \,\,=\frac{\mathrm{i}}{2}\left[\frac{\partial K^{\alpha\gamma}}{\partial p_{\mu}}\left(-p-k\right)e_{\mu}{}_{\gamma}^{\,\,\,\,\beta}-\frac{1}{2}\frac{\partial^{2}K^{\alpha\gamma}}{\partial p_{\mu}\partial p_{\nu}}\left(-p-k\right)k_{\mu}e_{\nu}{}_{\gamma}^{\,\,\,\,\beta}-\frac{\partial K^{\beta\gamma}}{\partial p_{\mu}}\left(p\right)e_{\mu}{}_{\gamma}^{\,\,\,\,\alpha}\right.\\
 & \,\,\,\,\,\,\,\left.-\frac{1}{2}\frac{\partial^{2}K^{\beta\gamma}}{\partial p_{\mu}\partial p_{\nu}}\left(p\right)k_{\mu}e_{\nu}{}_{\gamma}^{\,\,\,\,\alpha}\right],
\end{align*}
where I have absorbed the soft field polarization $e_{\mu}$ into
the definition of $\Gamma^{\left(3\right)}$. Notice that you need
to be careful about the order of $\alpha$ and $\beta$ when symmetrising.
We can simplify this to get, 
\begin{align}
 & \Gamma^{\left(3\right)}\left(e,k;p,-p-k\right)\nonumber \\
 & \,\,=\frac{\mathrm{i}}{2}\left[\frac{\partial K\left(-p-k\right)}{\partial p_{\mu}}e_{\mu}-\frac{1}{2}\frac{\partial^{2}K\left(-p-k\right)}{\partial p_{\mu}\partial p_{\nu}}k_{\mu}e_{\nu}-\left(\frac{\partial K\left(p\right)}{\partial p_{\mu}}e_{\mu}\right)^{T}-\frac{1}{2}\left(\frac{\partial^{2}K\left(p\right)}{\partial p_{\mu}\partial p_{\nu}}k_{\mu}e_{\nu}\right)^{T}\right]\nonumber \\
 & \,\,=\frac{\mathrm{i}}{2}\left[\frac{\partial K\left(-p\right)}{\partial p_{\mu}}e_{\mu}+\frac{1}{2}\frac{\partial^{2}K\left(-p\right)}{\partial p_{\mu}\partial p_{\nu}}k_{\mu}e_{\nu}-\left(\frac{\partial K\left(p\right)}{\partial p_{\mu}}e_{\mu}\right)^{T}-\frac{1}{2}\left(\frac{\partial^{2}K\left(p\right)}{\partial p_{\mu}\partial p_{\nu}}k_{\mu}e_{\nu}\right)^{T}\right]\nonumber \\
 & \,\,=\frac{\mathrm{i}}{2}\left[\frac{\partial K\left(-p\right)}{\partial p_{\mu}}e_{\mu}+\frac{1}{2}\frac{\partial^{2}K\left(-p\right)}{\partial p_{\mu}\partial p_{\nu}}k_{\mu}e_{\nu}-e_{\mu}^{T}\frac{\partial K^{T}\left(-p\right)}{\partial p_{\mu}}-\frac{1}{2}k_{\mu}e_{\nu}^{T}\frac{\partial^{2}K^{T}\left(-p\right)}{\partial p_{\mu}\partial p_{\nu}}\right]\nonumber \\
 & \,\,=\mathrm{i}\left[\frac{\partial K\left(-p\right)}{\partial p_{\mu}}e_{\mu}+\frac{1}{2}\frac{\partial^{2}K\left(-p\right)}{\partial p_{\mu}\partial p_{\nu}}k_{\mu}e_{\nu}\right],\label{eq:3-vertex}
\end{align}
where we've used (\ref{eq:symmetry of K}) in the second last line
of (\ref{eq:3-vertex}) and (\ref{eq:QK}) in the last line of (\ref{eq:3-vertex}).
If the soft boson line in $\Gamma^{\left(3\right)}$ is an internal
line, then (\ref{eq:3-vertex}) needs to be modified by stripping
of the polarization vector and contracting the index with the appropriate
soft propagator. Using (\ref{eq:3-vertex}) and the relations (\ref{eq:QK}),
(\ref{eq:onshell}), (\ref{eq:onshell1}) and (\ref{eq:onshell2}),
we can write-
\begin{align}
 & \Gamma^{\left(3\right)}\left(e,k;p,-p-k\right)\Xi\left(-p-k\right)\nonumber \\
 & \,\,=\mathrm{i}\left[\frac{\partial K\left(-p\right)}{\partial p_{\mu}}e_{\mu}+\frac{1}{2}\frac{\partial^{2}K\left(-p\right)}{\partial p_{\mu}\partial p_{\nu}}k_{\mu}e_{\nu}\right]\Xi\left(-p-k\right)\nonumber \\
 & \,\,=-\mathrm{i}\left[\frac{\partial K\left(-p\right)}{\partial p_{\mu}}\Xi\left(-p-k\right)+\frac{1}{2}\frac{\partial^{2}K\left(-p\right)}{\partial p_{\mu}\partial p_{\nu}}\Xi\left(-p-k\right)k_{\nu}\right]e_{\mu}^{T}\nonumber \\
 & \,\,=-\mathrm{i}\left[2\mathrm{i}p^{\mu}-K\left(-p\right)\frac{\partial\Xi\left(-p\right)}{\partial p_{\mu}}+\frac{\partial K\left(-p\right)}{\partial p_{\mu}}\frac{\partial\Xi\left(-p\right)}{\partial p_{\nu}}k_{\nu}+\frac{1}{2}\frac{\partial^{2}K\left(-p\right)}{\partial p_{\mu}\partial p_{\nu}}\Xi\left(-p\right)k_{\nu}\right]e_{\mu}^{T}\nonumber \\
 & \,\,=2p^{\mu}e_{\mu}^{T}+\mathrm{i}K\left(-p\right)\frac{\partial\Xi\left(-p\right)}{\partial p_{\mu}}e_{\mu}^{T}+\frac{\mathrm{i}}{2}\frac{\partial K\left(-p\right)}{\partial p_{\mu}}\frac{\partial\Xi\left(-p\right)}{\partial p_{\nu}}\left(k_{\mu}e_{\nu}-k_{\nu}e_{\mu}\right)^{T}\nonumber \\
 & \,\,\,\,\,\,\,+\frac{\mathrm{i}}{2}K\left(-p\right)\frac{\partial^{2}\Xi\left(-p\right)}{\partial p_{\mu}\partial p_{\nu}}k_{\nu}e_{\mu}^{T}.\label{eq: Gamma Xi}
\end{align}
 The Feynman rules for $A1$ give,
\begin{align}
A1= & \varepsilon_{i;\alpha}\Gamma^{\left(3\right)\alpha\gamma}\left(e,k;p_{i},-p_{i}-k\right)\frac{\Xi_{\gamma\beta_{i}}\left(-p_{i}-k\right)}{\left(-p_{i}-k\right)^{2}-m_{i}^{2}}G_{n}^{\left(i\right);\beta_{i}}\left(p_{i}+k\right)\nonumber \\
= & \frac{1}{p_{i}.k}\left[\varepsilon_{i;\alpha}p_{i}^{\mu}e_{\mu}^{T}{}_{\,\,\,\,\beta_{i}}^{\alpha}+\varepsilon_{i;\alpha}p_{i}^{\mu}e_{\mu}^{T}{}_{\,\,\,\,\beta_{i}}^{\alpha}k_{\rho}\frac{\partial}{\partial p_{i\rho}}\right.\nonumber \\
 & \left.+\frac{\mathrm{i}}{4}\varepsilon_{i;\alpha}\frac{\partial K^{\alpha\gamma}\left(-p_{i}\right)}{\partial p_{i\nu}}\frac{\partial\Xi_{\gamma\delta}\left(-p_{i}\right)}{\partial p_{i\mu}}\left(e_{\mu}^{T}k_{\nu}-e_{\nu}^{T}k_{\mu}\right)_{\,\,\,\,\beta_{i}}^{\delta}\right]G_{n}^{\left(i\right);\beta_{i}}\left(p_{i}\right)\nn\\
= & \frac{1}{p_{i}.k}\left[\varepsilon_{i}^{T}p_{i}^{\mu}e_{\mu}^{T}+k_{\rho}\varepsilon_{i}^{T}p_{i}^{\mu}e_{\mu}^{T}\frac{\partial}{\partial p_{i\rho}}+\frac{\mathrm{i}}{4}\varepsilon_{i}^{T}\frac{\partial K_{i}\left(-p_{i}\right)}{\partial p_{i\nu}}\frac{\partial\Xi_{i}\left(-p_{i}\right)}{\partial p_{i\mu}}\left(k_{\nu}e_{\mu}-k_{\mu}e_{\nu}\right)^{T}\right]G_{n}^{\left(i\right)}\left(p_{i}\right),\label{eq: A1}
\end{align}
where in the last line we have suppressed the representation indices
$\alpha$, $\beta$ etc. to tidy up the equation. As mentioned previously,
the notation $G_{n}^{\left(i\right),\alpha_{i}}$ is shorthand for
$\prod_{j\neq i}\varepsilon_{\alpha_{j}}G_{n}^{\alpha_{1}\ldots\alpha_{i}\ldots\alpha_{n}}$
which is the $n$ point Green's function with the $i$th polarization
vector amputated. The subscript $i$ in various places indicates that
the $i$th index is summed over in the implicit matrix product. Now
we want to evaluate the contribution due to diagrams of type $A2$.
For this we need the amputated Green's function $\tilde{G}_{n+1}\left(e,k;p_{1},\ldots,p_{n}\right)$
of $n$ finite energy fields and one soft gauge field which only includes
diagrams in which the soft particle is not attached to any external
leg. This is obtained by covariantising the amputated Green's function
$G_{n}^{\alpha_{1}\ldots\alpha_{n}}\left(p_{1},\ldots,p_{n}\right)$
of $n$ finite energy fields and taking only leading order contribution
(since the corresponding diagram is already subleading due to the
absence of a pole in the propagator). Hence we have,
\[
\tilde{G}_{n+1}\left(e,k;p_{1},\ldots,p_{n}\right)=-\sum_{i}\left(\prod_{j\neq i}\varepsilon_{\alpha_{j}}\right)\varepsilon_{\alpha_{i}}e_{\mu}{}_{\beta_{i}}^{\,\,\,\,\alpha_{i}}\frac{\partial}{\partial p_{i\mu}}G_{n}^{\alpha_{1}\ldots\beta_{i}\ldots\alpha_{n}}.
\]
The contribution due to the Feynman diagram of type $A2$ will be,
\[
A2=-\sum_{i}\varepsilon_{i}^{T}e_{\mu}\frac{\partial G_{n}^{\left(i\right)}}{\partial p_{j\mu}}.
\]
Summing over the soft boson attaching to all distinct external legs for
$A1$ and adding to $A2$ we get,
\begin{align}
G_{n+1}\left(e,k;\varepsilon_{i},p_{i}\right)= & \sum_{i}\varepsilon_{i}^{T}\frac{e^{T}.p_{i}}{p_{i}.k}G_{n}^{\left(i\right)}\left(p_{i}\right)\nonumber \\
 & +\sum_{i}\varepsilon_{i}^{T}\frac{e_{\mu}^{T}k_{\nu}}{p_{i}.k}\left\{ p_{i}^{\mu}\frac{\partial}{\partial p_{i\nu}}-p_{i}^{\nu}\frac{\partial}{\partial p_{i\mu}}\right\} G_{n}^{\left(i\right)}\left(p_{i}\right)\nonumber \\
 & +\sum_{i}\frac{1}{p_{i}.k}\frac{\mathrm{i}}{4}\varepsilon_{i}^{T}\frac{\partial K_{i}\left(-p_{i}\right)}{\partial p_{i\nu}}\frac{\partial\Xi_{i}\left(-p_{i}\right)}{\partial p_{i\mu}}\left(k_{\nu}e_{\mu}-k_{\mu}e_{\nu}\right)^{T}G_{n}^{\left(i\right)}\left(p_{i}\right).\label{eq:qed-soft-theorem-single}
\end{align}
Notice that the first two terms in (\ref{eq:qed-soft-theorem-single})
are independent of $K$ and $\Xi$, whereas the last term depends
on their derivatives and hence depends explicitly on the action.

\subsection*{Soft Photon Theorem in QED}

In the case of soft photons, the gauge group is simply $\text{U}\left(1\right)$
and the gauge field is Abelian. There is only one generator for the
lie algebra which is denoted by $Q_{\alpha}^{\,\,\,\,\beta}$ and
is sometimes called the charge matrix since its matrix elements are the electric
charges of the various fields $\phi_{\alpha}$ in the external state.
Therefore we have,
\begin{align}
G_{n+1}\left(e,k;\varepsilon_{i},p_{i}\right)= & \sum_{i}\varepsilon_{i}^{T}\frac{e.p_{i}}{p_{i}.k}Q^{T}G_{n}^{\left(i\right)}\left(p_{i}\right)\nonumber \\
 & +\sum_{i}\varepsilon_{i}^{T}Q^{T}\frac{e_{\mu}k_{\nu}}{p_{i}.k}\left\{ p_{i}^{\mu}\frac{\partial}{\partial p_{i\nu}}-p_{i}^{\nu}\frac{\partial}{\partial p_{i\mu}}\right\} G_{n}^{\left(i\right)}\left(p_{i}\right)\nonumber \\
 & +\sum_{i}\frac{1}{p_{i}.k}\frac{\mathrm{i}{\nonumber}}{4}\varepsilon_{i}^{T}\frac{\partial K_{i}\left(-p_{i}\right)}{\partial p_{i\nu}}\frac{\partial\Xi_{i}\left(-p_{i}\right)}{\partial p_{i\mu}}Q^{T}\left(k_{\nu}e_{\mu}-k_{\mu}e_{\nu}\right)G_{n}^{\left(i\right)}\left(p_{i}\right).\label{eq:qed-soft-theorem-single-1}
\end{align}
Now for the $i$th external particle of charge $q_{i}$
\[
Q\varepsilon_{i}=q_{i}\varepsilon_{i}.
\]
Therefore, 
\begin{align}
G_{n+1}\left(e,k;\varepsilon_{i},p_{i}\right)= & \sum_{i}\varepsilon_{i}^{T}q_{i}\frac{e.p_{i}}{p_{i}.k}G_{n}^{\left(i\right)}\left(p_{i}\right)\nonumber \\
 & +\sum_{i}\varepsilon_{i}^{T}q_{i}\frac{e_{\mu}k_{\nu}}{p_{i}.k}\left\{ p_{i}^{\mu}\frac{\partial}{\partial p_{i\nu}}-p_{i}^{\nu}\frac{\partial}{\partial p_{i\mu}}\right\} G_{n}^{\left(i\right)}\left(p_{i}\right)\nonumber \\
 & +\sum_{i}\frac{1}{p_{i}.k}\frac{\mathrm{i}}{4}\varepsilon_{i}^{T}q_{i}\frac{\partial K_{i}\left(-p_{i}\right)}{\partial p_{i\nu}}\frac{\partial\Xi_{i}\left(-p_{i}\right)}{\partial p_{i\mu}}\left(k_{\nu}e_{\mu}-k_{\mu}e_{\nu}\right)G_{n}^{\left(i\right)}\left(p_{i}\right).\label{eq:qed-soft-theorem-single-1-1}
\end{align}
If the $i$th external particle to which photon is attached is a Dirac
fermion,
\[
K_{i}\left(p_{i}\right)=\cancel{p_{i}}-m_{i},
\]
\[
\Xi_{i}\left(p_{i}\right)=\mathrm{i}\left(\cancel{p_{i}}+m_{i}\right),
\]
where $K_{i}$, $\Xi_{i}$ denote the $i$th blocks in $K$ and $\Xi$
corresponding to the $i$th particle. We have used the Feynman slash
notation to define $\cancel{p}=\gamma^{\mu}p_{\mu}$. For this case,
the theory dependent term is,
\begin{align*}
\frac{\mathrm{i}}{4}\left(e_{\mu}k_{\nu}-e_{\nu}k_{\mu}\right)q_{i}\varepsilon_{i}^{T}\frac{\partial K_{i}\left(-p_{i}\right)}{\partial p_{i\nu}}\frac{\partial\Xi_{i}\left(-p_{i}\right)}{\partial p_{i\mu}}= & -\frac{\mathrm{i}}{4}q_{i}\varepsilon_{i}^{T}\left(e_{\mu}k_{\nu}-e_{\nu}k_{\mu}\right)i\gamma^{\nu}\gamma^{\mu}\\
= & \frac{1}{4}q_{i}\varepsilon_{i}^{T}\left(\cancel{k}\cancel{e}-\cancel{e}\cancel{k}\right)\\
= & q_{i}\varepsilon_{i}^{T}\frac{1}{4}\left[\gamma^{\mu},\gamma^{\nu}\right]k_{\mu}e_{\nu}\\
= & -\mathrm{i}q_{i}\varepsilon_{i}^{T}\frac{k_{\mu}e_{\nu}}{p_{i}.k}S_{\left(i\right)}^{\mu\nu}.
\end{align*}
Lastly, for photons $\Gamma^{\left(3\right)}$ simply vanishes due
to charge conjugation invariance of the QED action (Furry's theorem).
Therefore the soft theorem in QED takes the form,
\begin{equation}
G_{n+1}\left(e,k;\varepsilon_{i},p_{i}\right)=\sum_{i}\varepsilon_{i}^{T}q_{i}\frac{e.p_{i}}{p_{i}.k}G_{n}^{\left(i\right)}\left(p_{i}\right)-\mathrm{i}\sum_{i}\varepsilon_{i}^{T}q_{i}\frac{e_{\mu}k_{\nu}}{p_{i}.k}\left(\hat{J}_{\left(i\right)}^{\mu\nu}G_{n}^{\left(i\right)}\right)\left(p_{i}\right),\label{eq:qed final soft theorem}
\end{equation}
where we have again identified the angular momentum operator $\hat{J}^{\mu\nu}$
acting on the Green's function $G_{n}^{\left(i\right)}$ as,
\begin{equation}
\left(\hat{J}_{\left(i\right)}^{\mu\nu}G_{n}^{\left(i\right)}\right)\left(p_{i}\right)=\mathrm{i}\left\{ p_{i}^{\mu}\frac{\partial}{\partial p_{i\nu}}-p_{i}^{\nu}\frac{\partial}{\partial p_{i\mu}}\right\} G_{n}^{\left(i\right)}\left(p_{i}\right)+\frac{\mathrm{i}}{4}\left[\gamma^{\mu},\gamma^{\nu}\right]G_{n}^{\left(i\right)}\left(p_{i}\right).
\end{equation}

\subsection*{Soft Gluon Theorem in QCD}

For gluons, the gauge group is $\text{SU}\left(3\right)$ and the Lie algebra
has 8 generators. If the external states are quarks, they will be
in the fundamental representation of $\text{SU}\left(3\right)$ in which
case the generators are given in terms of the 8 Gell-Mann matrices
$\lambda_{a}$,
\begin{equation}
T_{a}\varepsilon_{i}=g_{i}\frac{\lambda_{i;a}}{2}\varepsilon_{i},
\end{equation}
where $\lambda_{i;a}$ are such that the $i$th block is the corresponding
Gell-Mann matrix while the other blocks are identity. If the external
states are gluons then the generators are in the adjoint representation
of $\text{SU}\left(3\right)$,
\begin{equation}
T_{a}\varepsilon_{i}=g_{i}\tilde{T}_{i;a}\varepsilon_{i},
\end{equation}
where
\[
\left[\tilde{T}_{i;a}\right]_{bc}^{\mu\nu}=f_{abc}\eta^{\mu\nu}.
\]
In general we can write,
\[
T_{a}\varepsilon_{i}=g_{i}T_{i;a}\varepsilon_{i},
\]
In general, there can be some gluons and some quarks in the external
state. So we write the general soft gluon theorem as,
\begin{equation}
G_{n+1}\left(e,k;\varepsilon_{i},p_{i}\right)=\sum_{i}\varepsilon_{i}^{T}g_{i}\frac{e_{a}.p_{i}}{p_{i}.k}T_{i;a}G_{n}^{\left(i\right)}\left(p_{i}\right)-\mathrm{i}\sum_{i}\varepsilon_{i}^{T}g_{i}\frac{e_{a\mu}k_{\nu}}{p_{i}.k}T_{i;a}\left(\hat{J}_{\left(i\right)}^{\mu\nu}G_{n}^{\left(i\right)}\right)\left(p_{i}\right).\label{eq:qcd final soft theorem}
\end{equation}
This is the single soft gluon theorem in QCD. Note that the sum will only contain terms that are in accordance with the appropriate color ordering.

\section{Chern-Simons Vertices in D=4+1} \label{CS vertices}

Now that we have the subleading soft theorems we can move to the Chern-Simons
correction. For this we need to derive the tree-level Chern-Simons
vertices. Let us look at the case of Abelian gauge field first and
then the non-Abelian.

\subsection{Abelian Case: Chern-Simons QED}

Consider the Abelian gauge field $A_{\mu}$. The Chern-Simons
term in this case is given by,
\begin{align}
S_{\text{CS}} & =\kappa\int\mathbb{A}\left(d\mathbb{A}\right)^{2}=\kappa\int d^{4+1}x\,\epsilon^{\mu\nu\rho\sigma\delta}A_{\mu}\partial_{\nu}A_{\rho}\partial_{\sigma}A_{\delta}.
\end{align}
We can calculate the Chern-Simons
3-point vertex. The vertex factor in the Feynman rules is again found
by symmetrizing the Chern-Simons term in external momenta and factoring
out the external polarization factors $\varepsilon_{\mu}\left(k\right)$.
\begin{align}
\mathrm{i}\mathcal{M}= & \mathrm{i}\kappa\epsilon^{\mu\nu\rho\sigma\delta}\left[\varepsilon_{\mu}\left(k_{1}\right)\mathrm{i}k_{2\nu}\varepsilon_{\rho}\left(k_{2}\right)\mathrm{i}k_{3\sigma}\varepsilon_{\delta}\left(k_{3}\right)+\varepsilon_{\mu}\left(k_{1}\right)\mathrm{i}k_{3\nu}\varepsilon_{\rho}\left(k_{3}\right)\mathrm{i}k_{2\sigma}\varepsilon_{\delta}\left(k_{2}\right)\right.\nonumber \\
 & +\varepsilon_{\mu}\left(k_{2}\right)\mathrm{i}k_{3\nu}\varepsilon_{\rho}\left(k_{3}\right)\mathrm{i}k_{1\sigma}\varepsilon_{\delta}\left(k_{1}\right)+\varepsilon_{\mu}\left(k_{2}\right)\mathrm{i}k_{1\nu}\varepsilon_{\rho}\left(k_{1}\right)\mathrm{i}k_{3\sigma}\varepsilon_{\delta}\left(k_{3}\right)\nonumber \\
 & \left.+\varepsilon_{\mu}\left(k_{3}\right)\mathrm{i}k_{1\nu}\varepsilon_{\rho}\left(k_{1}\right)\mathrm{i}k_{2\sigma}\varepsilon_{\delta}\left(k_{2}\right)+\varepsilon_{\mu}\left(k_{3}\right)\mathrm{i}k_{2\nu}\varepsilon_{\rho}\left(k_{2}\right)\mathrm{i}k_{1\sigma}\varepsilon_{\delta}\left(k_{1}\right)\right]\nonumber \\
= & -2\mathrm{i}\,\kappa\epsilon^{\mu\nu\rho\sigma\delta}\left[k_{1\nu}k_{2\sigma}+k_{2\nu}k_{3\sigma}+k_{3\nu}k_{1\sigma}\right]\varepsilon_{\mu}\left(k_{1}\right)\varepsilon_{\rho}\left(k_{2}\right)\varepsilon_{\delta}\left(k_{3}\right).
\end{align}
Therefore, the vertex factor is-
\begin{equation}
\Gamma_{\text{CS}}^{\mu\rho\delta}\left(k_{1},k_{2},k_{3}\right)=-2\mathrm{i}\,\kappa\epsilon^{\mu\nu\rho\sigma\delta}\left[k_{1\nu}k_{2\sigma}+k_{2\nu}k_{3\sigma}+k_{3\nu}k_{1\sigma}\right]=-6\mathrm{i}\,\kappa\epsilon^{\mu\nu\rho\sigma\delta}k_{1\nu}k_{2\sigma}.\label{eq:CS vertex}
\end{equation}
The rest of the Feynman rules are same as those of usual QED.

\subsection{Non-Abelian Case: Chern-Simons QCD}

Now let us consider a non-Abelian gauge field $A_{\mu}$. 
\begin{equation}
S_{\text{CS}}=\kappa\int \text{Tr}\left[\mathbb{A}\left(d\mathbb{A}\right)^{2}+c_{4}\mathbb{A}^{3}d\mathbb{A}+c_{5}\mathbb{A}^{5}\right].
\end{equation}
As mentioned before, rational coefficients are fixed by gauge invariance. 

Using $S_{\text{CS}}$, the 3-point vertex is given by,
\begin{equation}
\Gamma_{\text{CS};abc}^{\left(3\right)\mu\rho\delta}\left(k_{1},k_{2},k_{3}\right)=-\frac{3\mathrm{i}}{2}\kappa\epsilon^{\mu\nu\rho\sigma\delta}d_{abc}k_{1\nu}k_{2\sigma},\label{eq:NA CS 3 vertex}
\end{equation}
where $\frac{\mathrm{i}}{2}d^{abc}=\text{Tr}\left[T^{a}\left\{ T^{b},T^{c}\right\} \right]$.
Similarly we can obtain the 4-point vertex to be,
\begin{align}
 & \Gamma_{\text{CS};abcd}^{\left(4\right)\mu\nu\rho\delta}\left(k_{1},k_{2},k_{3},k_{4}\right)\nonumber \\
 & \,\,=-c_{4}\kappa\epsilon^{\mu\nu\rho\sigma\delta}\left[k_{1\sigma}\left\{ f_{dce}\text{Tr}\left(T_{b}T_{e}T_{a}\right)+f_{bde}\text{Tr}\left(T_{c}T_{e}T_{a}\right)+f_{cbe}\text{Tr}\left(T_{d}T_{e}T_{a}\right)\right\} \right.\nonumber \\
 & \,\,\,\,\,\,\,+k_{2\sigma}\left\{ f_{cde}\text{Tr}\left(T_{a}T_{e}T_{b}\right)+f_{dae}\text{Tr}\left(T_{c}T_{e}T_{b}\right)+f_{ace}\text{Tr}\left(T_{d}T_{e}T_{b}\right)\right\} \nonumber \\
 & \,\,\,\,\,\,\,+k_{3\sigma}\left\{ f_{dbe}\text{Tr}\left(T_{a}T_{e}T_{c}\right)+f_{ade}\text{Tr}\left(T_{b}T_{e}T_{c}\right)+f_{bae}\text{Tr}\left(T_{d}T_{e}T_{c}\right)\right\} \nonumber \\
 & \,\,\,\,\,\,\,\left.+k_{4\sigma}\left\{ f_{bce}\text{Tr}\left(T_{a}T_{e}T_{d}\right)+f_{cae}\text{Tr}\left(T_{b}T_{e}T_{d}\right)+f_{abe}\text{Tr}\left(T_{c}T_{e}T_{d}\right)\right\} \right].
\end{align}
Using the definitions of $f_{abc}$ and $d_{abc}$ and simplifying
using the Jacobi identity gives,
\begin{align}
 & \Gamma_{\text{CS};abcd}^{\left(4\right)\mu\nu\rho\delta}\left(k_{1},k_{2},k_{3},k_{4}\right)\nonumber \\
 & \,\,=-\frac{c_{4}}{4}\mathrm{i}\kappa\epsilon^{\mu\nu\rho\sigma\delta}\left[\left(k_{1\sigma}-k_{2\sigma}\right)f_{dce}d_{bea}+\left(k_{1\sigma}-k_{3\sigma}\right)f_{bde}d_{cea}+\left(k_{1\sigma}-k_{4\sigma}\right)f_{cbe}d_{dea}\right.\nonumber \\
 & \,\,\,\,\,\,\,\left.+\left(k_{2\sigma}-k_{3\sigma}\right)f_{dae}d_{ceb}+\left(k_{2\sigma}-k_{4\sigma}\right)f_{ace}d_{deb}+\left(k_{3\sigma}-k_{4\sigma}\right)f_{bae}d_{dec}\right].
\end{align}
We can also calculate the 5-point vertex $\Gamma_{\text{CS}}^{\left(5\right)}$
but for our purpose it will suffice to just note that it will have
no momentum dependence at tree-level.

\section{Chern-Simons Correction to Single Soft Photon Theorem in QED} \label{CS correction QED}

Now we add the Chern-Simons term to the action and see how it modifies
the soft photon theorem. We start with considering a single soft photon
in the external state. Then the relevant diagrams are the ones of
type $A1$ and $A2$. From (\ref{eq:CS vertex}), we see that the
Chern-Simons vertex modifies $\Gamma^{\left(3\right)}$ only when
the external hard particle is the massless $\text{U}\left(1\right)$ gauge
boson. In that case we have,
\begin{align}
\delta\Gamma^{\left(3\right)\alpha\beta}\left(e,k;p,-p-k\right)= & e_{\mu}\Gamma_{\text{CS}}^{\mu\alpha\beta}\left(k,-p-k,p\right)\nonumber \\
= & 6\mathrm{i}\kappa\epsilon^{\mu\nu\alpha\sigma\beta}e_{\mu}k_{\nu}p_{\sigma},
\end{align}
where in this case the representation indices $\alpha$, $\beta$,
$\gamma$ are the same as the Lorentz indices since the external line
to which the soft photon attaches is also a photon line (with finite
energy). . For a finite energy photon field,
\[
K_{\text{photon}}^{\mu\nu}\left(p\right)=-p^{2}\eta^{\mu\nu},
\]
\[
\Xi_{\mu\nu}^{\text{photon}}\left(p\right)=-\mathrm{i}\eta_{\mu\nu}.
\]
Hence, $A1$ will receive a correction (at the subleading level) of
the form,
\begin{align*}
\delta A1= & \varepsilon^{T}\left(p_{i}\right)\delta\Gamma^{\left(3\right)}\left(e,k;p_{i},-p_{i}-k\right)\frac{\Xi_{i}^{\text{photon}}\left(-p_{i}-k\right)}{2p_{i}.k}G_{n}^{\left(i\right)}\left(p_{i}+k\right)\\
= & 3\mathrm{i}\kappa\frac{1}{p_{i}.k}\varepsilon_{i;\alpha}\epsilon^{\mu\nu\alpha\sigma\beta}e_{\mu}k_{\nu}p_{i\sigma}\Xi_{\beta\gamma}^{\text{photon}}\left(-p_{i}\right)G_{n}^{\left(i\right);\gamma}\left(p_{i}\right)\\
= & 3\kappa\epsilon^{\mu\nu\alpha\sigma\beta}\frac{e_{\mu}k_{\nu}}{p_{i}.k}\varepsilon_{i;\alpha}p_{i\sigma}\eta_{\beta\gamma}G_{n}^{\left(i\right);\gamma}\left(p_{i}\right).
\end{align*}
Since, the Chern-Simons vertex is already subleading, the Chern-Simons
correction to $A2$ is higher order. Hence for single soft photon
emission, the Chern-Simons modifies (\ref{eq:qed-soft-theorem-single})
to give,
\begin{align}
G_{n+1}\left(e,k;\varepsilon_{i},p_{i}\right)= & \sum_{i}\frac{1}{p_{i}.k}\varepsilon_{i}^{T}e_{\mu}p_{i}^{\mu}Q_{i}^{T}G_{n}^{\left(i\right)}\left(p_{i}\right)\nonumber \\
 & +\sum_{i}\frac{1}{p_{i}.k}e_{\mu}k_{\nu}\varepsilon_{i}^{T}Q_{i}^{T}\left\{ p_{i}^{\mu}\frac{\partial}{\partial p_{i\nu}}-p_{i}^{\nu}\frac{\partial}{\partial p_{i\mu}}\right\} G_{n}^{\left(i\right)}\left(p_{i}\right)\nonumber \\
 & +\sum_{i}\frac{1}{p_{i}.k}\frac{\mathrm{i}}{4}\left(e_{\mu}k_{\nu}-e_{\nu}k_{\mu}\right)\varepsilon_{i}^{T}\frac{\partial K_{i}\left(-p_{i}\right)}{\partial p_{i\nu}}\frac{\partial\Xi_{i}\left(-p_{i}\right)}{\partial p_{i\mu}}Q_{i}^{T}G_{n}^{\left(i\right)}\left(p_{i}\right)\nonumber \\
 & +\widetilde{\sum_{j}}3\kappa\epsilon^{\mu\nu\alpha\sigma\beta}\frac{e_{\mu}k_{\nu}}{p_{j}.k}\varepsilon_{j;\alpha}p_{j\sigma}G_{n;\beta}^{\left(j\right)}\left(p_{j}\right),\label{eq:CS-soft-theorem-single}
\end{align}
where $\widetilde{\Sigma}$ only sums over external photon legs.

\section{Chern-Simons Correction to Single Soft Gluon Theorem in QCD} \label{CS correction QCD}

For single soft gluon emission, the story does not change much. Again
$A2$ is subleading and the correction to $A1$ is given by ,
\begin{align}
\delta\Gamma_{ab}^{\left(3\right)\alpha\beta}\left(e,k;p,-p-k\right)= & e_{c\mu}\Gamma_{\text{CS};cab}^{\mu\alpha\beta}\left(k,-p-k,p\right)\nonumber \\
= & \frac{3\mathrm{i}\kappa}{2}\epsilon^{\mu\nu\alpha\sigma\beta}d_{cab}e_{c\mu}k_{\nu}p_{\sigma},\label{eq:CS correction NA}
\end{align}
where in this case the representation indices are the pair $\left\{ a,\alpha\right\} $
of color indices and Lorentz indices since the external line to which
the soft gluon attaches is also a gluon line (with finite energy).
Also,
\begin{equation}
K_{ab}^{\text{gluon};\mu\nu}\left(p\right)=-p^{2}\eta^{\mu\nu}\delta_{ab},\label{eq:K gluon}
\end{equation}
\begin{equation}
\Xi_{\mu\nu ab}^{\text{gluon}}\left(p\right)=-\mathrm{i}\eta_{\mu\nu}\delta_{ab}.\label{eq:Xi gluon}
\end{equation}
Hence, $A1$ will receive a correction (at the subleading level) of
the form,
\begin{align*}
\delta A1= & \varepsilon^{T}\left(p_{i}\right)\delta\Gamma^{\left(3\right)}\left(e,k;p_{i},-p_{i}-k\right)\frac{\Xi_{i}\left(-p_{i}-k\right)}{2p_{i}.k}G_{n}^{\left(i\right)}\left(p_{i}+k\right)\\
= & \frac{3\mathrm{i}\kappa}{4}\frac{1}{p_{i}.k}\varepsilon_{i;a\alpha}\epsilon^{\mu\nu\alpha\sigma\beta}d_{cab}e_{c\mu}k_{\nu}p_{i\sigma}\Xi_{\beta\gamma bd}^{\text{gluon}}\left(-p_{i}\right)G_{n;d}^{\left(i\right);\gamma}\left(p_{i}\right)\\
= & \frac{3\kappa}{4}\epsilon^{\mu\nu\alpha\sigma\beta}d_{cab}\frac{e_{c\mu}k_{\nu}}{p_{i}.k}\varepsilon_{i;a\alpha}p_{i\sigma}\eta_{\beta\gamma}G_{n;b}^{\left(i\right);\gamma}\left(p_{i}\right).
\end{align*}
Since, the Chern-Simons vertex is already subleading, the Chern-Simons
correction to $A2$ is higher order. Hence for single soft gluon emission,
the Chern-Simons modifies (\ref{eq:qed-soft-theorem-single}) to give,
\begin{align}
G_{n+1}\left(e,k;\varepsilon_{i},p_{i}\right)= & \sum_{i}\varepsilon_{i}^{T}g_{i}\frac{e_{a}.p_{i}}{p_{i}.k}T_{i;a}G_{n}^{\left(i\right)}\left(p_{i}\right)\nonumber \\
 & -\mathrm{i}\sum_{i}\varepsilon_{i}^{T}g_{i}\frac{e_{a\mu}k_{\nu}}{p_{i}.k}T_{i;a}\left(\hat{J}_{\left(i\right)}^{\mu\nu}G_{n}^{\left(i\right)}\right)\left(p_{i}\right)\nonumber \\
 & +\widetilde{\sum_{j}}\frac{3\kappa}{4}\epsilon^{\mu\nu\alpha\sigma\beta}d_{cab}\frac{e_{c\mu}k_{\nu}}{p_{j}.k}\varepsilon_{j;a\alpha}p_{j\sigma}G_{n;b\beta}^{\left(j\right)}\left(p_{j}\right),\label{eq: CS QCD}
\end{align}
where $\widetilde{\Sigma}$ only sums over external gluon legs. Again, we assume that the sums only contain terms that are in accordance with the color ordering of the amplitude.

\section{Chern-Simons Correction to Soft Theorems with Multiple Simultaneous
Soft Emissions} \label{CS correction multiple}

\begin{figure}
\centering

\tikzset{every picture/.style={line width=0.75pt}} 

\begin{tikzpicture}[x=0.75pt,y=0.75pt,yscale=-1,xscale=1]

\draw    (103,162) -- (142,162) ;
\draw   (181,162) .. controls (181,148.19) and (192.19,137) .. (206,137) .. controls (219.81,137) and (231,148.19) .. (231,162) .. controls (231,175.81) and (219.81,187) .. (206,187) .. controls (192.19,187) and (181,175.81) .. (181,162) -- cycle ;
\draw   (309,162) .. controls (309,123.06) and (340.56,91.5) .. (379.5,91.5) .. controls (418.44,91.5) and (450,123.06) .. (450,162) .. controls (450,200.94) and (418.44,232.5) .. (379.5,232.5) .. controls (340.56,232.5) and (309,200.94) .. (309,162) -- cycle ;
\draw    (142,162) -- (181,162) ;
\draw [shift={(144,162)}, rotate = 180] [fill={rgb, 255:red, 0; green, 0; blue, 0 }  ][line width=0.08]  [draw opacity=0] (12,-3) -- (0,0) -- (12,3) -- cycle    ;
\draw    (231,162) -- (268,162) ;
\draw [shift={(270,162)}, rotate = 180] [fill={rgb, 255:red, 0; green, 0; blue, 0 }  ][line width=0.08]  [draw opacity=0] (12,-3) -- (0,0) -- (12,3) -- cycle    ;
\draw    (270,162) -- (309,162) ;
\draw    (420,104) -- (453,75) ;
\draw [shift={(451.5,76.32)}, rotate = 318.69] [fill={rgb, 255:red, 0; green, 0; blue, 0 }  ][line width=0.08]  [draw opacity=0] (12,-3) -- (0,0) -- (12,3) -- cycle    ;
\draw    (453,75) -- (484,47) ;
\draw    (449,150) -- (494,137) ;
\draw [shift={(492.08,137.56)}, rotate = 343.89] [fill={rgb, 255:red, 0; green, 0; blue, 0 }  ][line width=0.08]  [draw opacity=0] (12,-3) -- (0,0) -- (12,3) -- cycle    ;
\draw    (494,137) -- (545,123) ;
\draw    (428,214) -- (456,238) ;
\draw [shift={(454.48,236.7)}, rotate = 40.6] [fill={rgb, 255:red, 0; green, 0; blue, 0 }  ][line width=0.08]  [draw opacity=0] (12,-3) -- (0,0) -- (12,3) -- cycle    ;
\draw    (456,238) -- (499,275) ;
\draw  [dash pattern={on 0.84pt off 2.51pt}]  (206,187) -- (206,203) ;
\draw  [dash pattern={on 0.84pt off 2.51pt}]  (206.06,205) -- (207,236) ;
\draw [shift={(206,203)}, rotate = 88.26] [fill={rgb, 255:red, 0; green, 0; blue, 0 }  ][line width=0.08]  [draw opacity=0] (12,-3) -- (0,0) -- (12,3) -- cycle    ;
\draw    (449,169) -- (496,184) ;
\draw [shift={(494.09,183.39)}, rotate = 17.7] [fill={rgb, 255:red, 0; green, 0; blue, 0 }  ][line width=0.08]  [draw opacity=0] (12,-3) -- (0,0) -- (12,3) -- cycle    ;
\draw    (496,184) -- (548,199) ;
\draw   (182,261) .. controls (182,247.19) and (193.19,236) .. (207,236) .. controls (220.81,236) and (232,247.19) .. (232,261) .. controls (232,274.81) and (220.81,286) .. (207,286) .. controls (193.19,286) and (182,274.81) .. (182,261) -- cycle ;
\draw  [dash pattern={on 0.84pt off 2.51pt}]  (190,279) -- (181,300) ;
\draw  [dash pattern={on 0.84pt off 2.51pt}]  (180.3,301.87) -- (169,332) ;
\draw [shift={(181,300)}, rotate = 110.56] [fill={rgb, 255:red, 0; green, 0; blue, 0 }  ][line width=0.08]  [draw opacity=0] (12,-3) -- (0,0) -- (12,3) -- cycle    ;
\draw  [dash pattern={on 0.84pt off 2.51pt}]  (224,280) -- (236,304) ;
\draw  [dash pattern={on 0.84pt off 2.51pt}]  (236.85,305.81) -- (251,336) ;
\draw [shift={(236,304)}, rotate = 64.89] [fill={rgb, 255:red, 0; green, 0; blue, 0 }  ][line width=0.08]  [draw opacity=0] (12,-3) -- (0,0) -- (12,3) -- cycle    ;

\draw (182,169) node [anchor=north west][inner sep=0.75pt]   [align=left] {$ $};
\draw (194,154) node [anchor=north west][inner sep=0.75pt]   [align=left] {$\displaystyle \Gamma ^{( 3)}$};
\draw (357,139) node [anchor=north west][inner sep=0.75pt]  [font=\LARGE] [align=left] {$\displaystyle G_{n}^{( i)}$};
\draw (130,136) node [anchor=north west][inner sep=0.75pt]   [align=left] {$\displaystyle \varepsilon _{i}$,$\displaystyle p_{i}$};
\draw (223,132) node [anchor=north west][inner sep=0.75pt]  [font=\scriptsize] [align=left] {{\scriptsize $\displaystyle p_{i} +k_{1} +...k_{4}$}};
\draw (221,200) node [anchor=north west][inner sep=0.75pt]   [align=left] {$\displaystyle k_{1} +...k_{4}$};
\draw (409,57) node [anchor=north west][inner sep=0.75pt]   [align=left] {$\displaystyle \varepsilon _{1}$,$\displaystyle p_{1}$};
\draw (447,96) node [anchor=north west][inner sep=0.75pt]   [align=left] {$\displaystyle ...\varepsilon _{i-1}$,$\displaystyle p_{i-1}$};
\draw (451,198) node [anchor=north west][inner sep=0.75pt]   [align=left] {$\displaystyle \varepsilon _{i+1}$,$\displaystyle p_{i+1}$...};
\draw (417,247) node [anchor=north west][inner sep=0.75pt]   [align=left] {$\displaystyle \varepsilon _{n}$,$\displaystyle p_{n}$};
\draw (249,287) node [anchor=north west][inner sep=0.75pt]   [align=left] {$\displaystyle e_{4}$,$\displaystyle k_{4}$};
\draw (137,290) node [anchor=north west][inner sep=0.75pt]   [align=left] {$\displaystyle e_{1}$,$\displaystyle k_{1}$};
\draw (194,252) node [anchor=north west][inner sep=0.75pt]   [align=left] {$\displaystyle \Gamma _{\text{CS}}^{( 5)}$};
\draw (183,303) node [anchor=north west][inner sep=0.75pt]   [align=left] {.};
\draw (201,308) node [anchor=north west][inner sep=0.75pt]   [align=left] {.};
\draw (219,306) node [anchor=north west][inner sep=0.75pt]   [align=left] {.};
\draw (191,306) node [anchor=north west][inner sep=0.75pt]   [align=left] {.};
\draw (225,302) node [anchor=north west][inner sep=0.75pt]   [align=left] {.};
\draw (210,308) node [anchor=north west][inner sep=0.75pt]   [align=left] {.};

\end{tikzpicture}
\caption{Diagram containing a 5-point Chern-Simons vertex contributing to the
subleading soft theorem with four soft emissions\label{fig:5P}}
\end{figure}
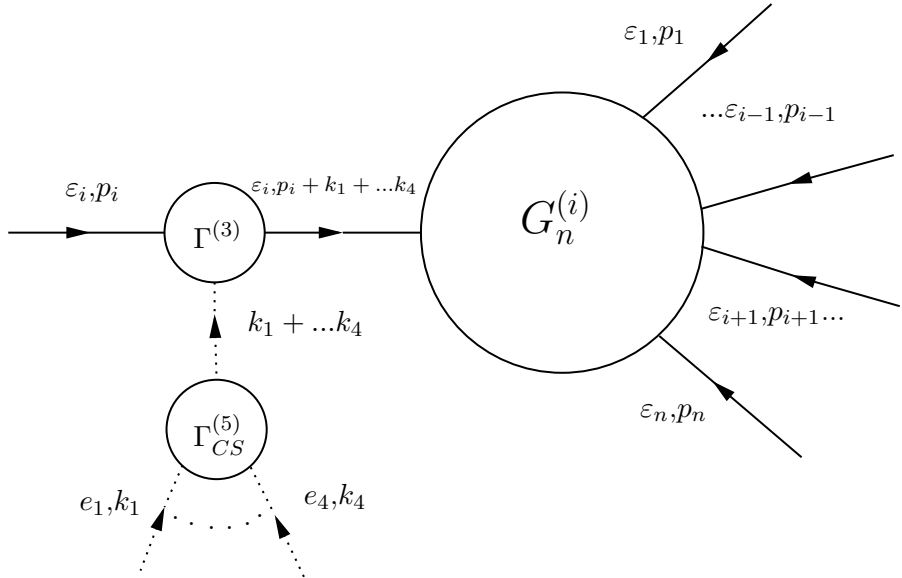

We now want to consider multiple soft boson emissions where the energies
of the emitted soft particles are of the same order, i.e. the soft
limits of amplitudes are taken simultaneously for all soft bosons.
One important thing to note is that all the Chern-Simons vertices
have one extra power of momentum and therefore are subleading in the
small $k$ limit to their usual gauge theory counterparts,
\[
\Gamma_{\text{CS}}^{\left(3\right)}\sim k\,\Gamma_{\text{gauge}}^{\left(3\right)},
\]
\[
\Gamma_{\text{CS}}^{\left(4\right)}\sim k\,\Gamma_{\text{gauge}}^{\left(4\right)}.
\]
This implies that there are no diagrams involving these vertices that
contribute to the leading soft theorem. This can be seen by realizing
that if there were any such diagrams we could have replaced the Chern-Simons
vertices with their pure gauge counterparts and would have gotten
an even more dominant contribution. This also implies that the Chern-Simons
corrections to subleading soft theorems come largely from adding the
Chern-Simons correction to the gauge vertices in the diagrams contributing
to the leading soft theorem, as was seen in the case of a single soft
emission. The only exception are diagrams that have the 5-point Chern-Simons
vertex as they do not have a corresponding more dominant gauge vertex.
But we can argue that these also lead to only subleading contributions.
For instance, consider figure \ref{fig:5P} which contributes to the
Chern-Simons correction for the subleading soft theorem with four
soft emissions. This is the only diagram containing the 5-point vertex
which contributes at the subleading order for this process, i.e. at
$\mathcal{O}\left(\nicefrac{1}{k^{3}}\right)$. It is easy to see
that there is no diagram containing the 5-point vertex which is more
dominant than this since it maximizes the negative $k$ powers in
the propagator while having the 5-point Chern-Simons vertex.
 Similar argument holds for soft theorems with even more soft emissions.
Hence, we can safely conclude that even for multiple soft emissions,
the Chern-Simons corrections to soft theorem for tree-level amplitudes are subleading to the gauge theory.

\section{Chern-Simons Terms in Higher Dimensions}

Finally let us comment on what happens if you add a perturbative Chern-Simons terms to the gauge theory in space-time dimensions other than five. As we stated earlier, in $D=2+1$ the Chern-Simons term is more dominant than the gauge term in the infrared. Therefore, we cannot treat the Chern-Simons term as a perturbation over the gauge theory in three dimensions. In $D>4+1$, the 3-point Chern-Simons vertex is absent\footnote{For instance, in $D=6+1$ the lowest vertex is the 4-point vertex which goes as $\sim\int\mathbb{A}d\mathbb{A}^3$}. Therefore, it is expected that the Chern-Simons term will not correct the subleading soft theorem at the subleading order in $D>4+1$.  Therefore, one concludes that the universality of the subleading soft theorem is still maintained in $D>4+1$.

\section{Conclusion} \label{conclusion}

We have shown in (\ref{eq:CS-soft-theorem-single}) and (\ref{eq: CS QCD})
that the Chern-Simons corrections to the soft photon and soft gluon
theorems are subleading in powers of soft momenta $k$. The leading soft theorems remain unchanged even in the presence of a Chern-Simons term. This is interesting for two reasons -

(i) the subleading term gets modified when Chern-Simons terms are added and therefore, it is no longer universal even though the amplitude is still gauge-invariant under small gauge transformations. Also, such terms cannot be obtained from Sen's covariantisation procedure and need to be added separately.

(ii) we can in principle take arbitrary values of $\left\{ c_{4},c_{5}\right\}$ which would leave the deformation $S_{CS}$ completely non-gauge-invariant. Even under such non-gauge-invariant deformations, the amplitudes still obey the "universal" leading soft theorem, at least at the tree-level. Therefore, we have an example of a non-gauge-invariant action that still satisfies the usual universal leading soft theorem at the tree-level. This means that gauge invariance of amplitudes is a sufficient but not necessary condition for the universal form of the tree-level leading soft theorem.

The analysis in this paper is limited to tree-level amplitudes. Calculations at higher loop orders may modify the second statement since the loss of gauge invariance should imply fewer constraints on the amplitudes. Therefore, one would expect that to show up in the IR behaviour of the amplitudes at higher loop orders.

\section*{Acknowledgments}
I would like to thank, R. Loganayagam, for helping
me with the initial idea and in setting up the problem as well as
for countless valuable insights. I would also like to thank Biswajit Sahoo and Ashoke Sen for their invaluable feedback on the draft. I would also like to thank the referee for their insightful comments on the manuscript. Finally,
I would like to thank my colleagues Godwin Martin, Kaustubh Singhi
and Shridhar Vinayak for contributing to various discussions throughout
this work. I also acknowledge the support of the Department of Atomic
Energy, Government of India, project no. RTI4019.

\bibliographystyle{jhepbibstyle}
\bibliography{references}
\end{document}